\newcommand{\Pl}{\partial}
\newcommand{\ts}{\textstyle}
\newcommand{\fder}[2]{\frac{{\ts d \/ #1}}{{\ts d\/ #2}}}
\newcommand{\fpar}[2]{\frac{{\ts \Pl \/ #1}}{{\ts \Pl \/ #2}}}

\newcommand{\bee}{\begin{equation}}
\newcommand{\ene}{\end{equation}}
\newcommand{\beea}{\begin{eqnarray}}
\newcommand{\enea}{\end{eqnarray}}
\documentclass[prb]{revtex4}
\usepackage{dcolumn}
\usepackage{graphicx}
\usepackage{latexsym}
\usepackage{amsmath}
\usepackage{amssymb}
\begin{document}
\title{Viscoelastic modes in a strongly coupled cold magnetized dusty plasma}
\author{Debabrata Banerjee,  Janaki Sita Mylavarapu and Nikhil Chakrabarti}
\affiliation{ Saha Institute of Nuclear Physics, I/AF Bidhannagar,
Calcutta 700 064, India}

        \begin{abstract}
        A generalized hydrodynamical model has been used to study low frequency modes in
        a strongly coupled, cold, magnetized dusty plasma. Such plasmas exhibit
        elastic properties due to strong correlations among dust
        particles and the tensile stresses imparted by the magnetic field.
        It has been shown that longitudinal compressional Alfven modes and
        elasticity modified transverse shear mode exist  in such a medium.
        The features of these collective modes are established and discussed.

        \end{abstract}
\maketitle

\section{Introduction}
Dusty plasmas are electron ion plasmas together with micron-sized
dust grains that can carry several thousand elementary charges.
The competition between the average Coulomb interaction energy
between the dust particles and the average thermal energy is
characterized by the Coulomb coupling parameter $\Gamma =
q_d^2/(k_B T_d\; r) $ where $q_d$ is the charge on the dust
grains, $r(\simeq n_d^{-1/3})$ is the average distance between
them, $T_d$ is the temperature of the dust component and $k_B$ is
the Boltzmann constant. The high, typically negative charge on the
dust leads to large values of the Coulomb parameter $\Gamma \gg1$
even at room temperature and the plasma is said to be in a
strongly coupled state\cite{shma}. A broad variety of systems in the
astrophysical context such as interior of heavy planets, white
dwarfs, neutron stars have matter in the strongly coupled state\cite{verh}.
In  the laboratories, we have examples of strongly coupled plasmas
produced for plasma processing and industrial applications, in
semiconductor heterojunctions and in the laser implosion
experiments with $\Gamma$ values in the range $1-100$. Both
crystalline and fluid properties can coexist in a strongly coupled
dusty plasma \cite{ikej} (or colloidal
plasma). While in such a state, a plasma possess viscous
properties typical of fluids as well as elastic properties similar
to solids \cite{pkaw}.  The dependence of viscous and elastic coefficients of
a medium on $\Gamma$ is  known in the case of  one component
plasmas and Yukawa liquids. The well known weakly coupled ideal
coulomb plasma (gas phase) characterized by $\Gamma \ll1$  has no
elastic property and trivial role of viscosity. At $\Gamma\sim1$
viscosity comes into the system profoundly and as $\Gamma$
increases elastic property gradually becomes important as the
plasma state switches over from the gaseous phase to liquid phase.
When $\Gamma>\Gamma_{c}$ (beyond $\Gamma_{c}$ system becomes
crystalline), viscosity disappears and  only elasticity reigns
over the system. So in the regime of $\Gamma$ from $1$ to
$\Gamma_{c}$ both viscosity and elasticity are of  simultaneous
concern and this property together is known as visco-elasticity.
An analogous behavior occurs in the case of fluids made of large
macromolecules such as polymer molecules that also exhibit
viscoelastic property in contrast to fluids made of small
molecules\cite{ogil}.

The presence of dust grains in a plasma with their charges and
masses that are orders of magnitudes higher than that of ions give
rise to new wave phenomena that are associated with longer time
and length scales. In the strongly coupled state, the dusty plasma
offers yet another advantage of studying wave phenomena that are
typical of  solids such as the transverse waves.

Using generalized hydrodynamic model\cite{pkaw,ichi}, many authors had shown that
longitudinal dust acoustic mode must be  corrected by an
additional term in the strongly coupled regime. They also found a
new transverse shear-like mode which comes into effect due to the
elastic property imparted to the medium due to the presence of
dust particles.
In a weakly coupled magnetized plasma, it is well known that the
presence of magnetic field makes the medium elastic, enabling
propagation of shear Alfven waves  with Alfven speed
$V_A^2=B^2/\mu_0\rho_0$.  Such kind of transverse waves do not
exist in an un-magnetized plasma in the weakly coupled state.  On
the other hand, the presence of strong correlations among
particles, leads to mechanical shear stresses that sustain the
propagation of transverse waves like in solids. It is of interest
to see the nature of the  elastic mode that can propagate in a
cold magnetized strongly coupled plasma\cite{sami}. In this mode the effects of the
elasticity due to the simultaneous presence of magnetic field as
well as due to strongly coupled nature of the dust particles  are
taken into account. Basic equations for this study has been written in the framework
of a generalized  magnetohydrodynamic
model to include viscoelastic effects.  It is found
that for both longitudinal and transverse  perturbations,
elasticity modified Alfven type modes can propagate  that can be
termed `magnetoelastic modes'. The dynamics of the outer crust of
magnetized neutron stars consisting of elastic solid media
permeated by frozen magnetic fields is likely to be governed by
such modes.

In Section -II we present the generalized hydrodynamic model containing the  Basic equations
which supports viscoelastic stresses.    In section III, we describe the analysis of
mode dispersion  by a linear stability analysis. In section IV, we have presented a short
summary of this work.

\section{Model and Basic Equations}
 We shall assume that the strongly coupled dusty plasma consists of the
electrons and ions that
 are weakly coupled and highly charged dust particles with strong
correlations among them \cite{pks}.
 We also consider the effects of an external magnetic field ${\bf B} = B_0
\hat z $
 on such a strongly coupled plasma. The medium acquires an elastic
property  because of the
  tensile stresses exerted by the magnetic field lines  as well as the
strongly correlated dust grains.     We wish to write down the
magnetohydrodynamic equations  for a strongly coupled magnetized dusty
plasma with an intention to study the dispersion relations which exhibit
coupling between  low frequency waves
  that arise due to  magnetic  and solid-like stresses.
  We assume the characteristic wave frequency to be much smaller than the
ion gyrofrequency,
  where dust dynamics is important. In such a situation
  the ion and electron inertial forces are much smaller than the
corresponding
 Lorentz forces. Therefore equations of motion for the electron
 and ion fluids can be written as:
 \begin{equation}
 0= - en_{e}({\bf E} + {{\bf v}_{e}\times {\bf B}}),
 \label{eem}
 \end{equation}
 \begin{equation}
0 =  en_{i}({\bf E} + {{\bf v}_{i} \times {\bf B}}),
 \label{iem}
\end{equation}
where, $n_{e,i} $ is the number density  of electrons and ions
fluid and ${\bf v}_{e,i}$ is corresponding velocity. The electric
and magnetic fields are   denoted as ${\bf E}$ and ${\bf B}$
respectively.
 For the dust fluid we adopt the generalized magnetohydrodynamic
 model\cite{mamu}. Since a micron size  dust grain may contain
several thousand elementary
 charges, in principle dust fluid is highly viscous compared to electron
and ion fluids,
 so we consider  viscosity terms only in the context of the dust momentum equation.
The momentum equation for the dust fluid  is
\begin{equation}
 m_d n_d\frac{d {\bf v}_{d} }{dt} =  Z en_{d}({\bf E} + {{\bf v}_{d}
\times {\bf B}})
  +\eta \bigtriangledown^{2}{\bf v}_{d}+
 \left (\xi+\frac{\eta}{3}\right )\nabla (\nabla.{\bf v}_d),
\label{dme}
\end{equation}
where, $n_{d} $ is the number density of dust fluid,  ${\bf v}_d$
is the dust fluid velocity, $Z$ is the number of negative charges
on a single dust particle, $\eta$ and $\xi$ are the dust shear viscosity and
 bulk viscosity coefficient respectively. The physical
interpretation of the above equation is illustrated in  Ref.
\cite{pkaw}.

 Next we shall define the mass
density, center of mass fluid flow velocity and current density
for the bulk fluid. First mass density may be defined as $\rho_m=
m_e n_e+m_in_i +m_d n_d$. Since $m_e,m_i \ll m_d$, $ \rho_m=
\rho_d \approx m_d n_d $. Then bulk velocity
 ${\bf v} = (m_e n_e {\bf v}_e +
m_i n_i{\bf v}_e + m_d n_d {\bf v}_d)/(m_d n_d) \approx
{\bf v}_{d}$ and finally the current density is defined as $
{\bf J} = e(n_i{\bf v}_i-n_e {\bf v}_e -Z n_d {\bf v}_d)$.
The current density ${\bf J}$ related to the magnetic field ${\bf
B}$ through Ampere's law is given by
 \begin{equation}
 \nabla\times {\bf B}= \mu_0 {\bf J}
  =\mu_0 e(n_{i}{\bf v}_{i}-n_{e}{\bf v}_{e}-Zn_{d} {\bf v}_d).
  \label{al}
  \end{equation}
 Note here that we have neglected the displacement current
 in the above equation since we are interested in low frequency
 ($\omega \ll \omega_{pd}$) and long wavelength pertubations. By adding the
 equations (\ref{eem}),
(\ref{iem}) and
 (\ref{dme}) and working in the MHD
 approximation for an viscous dusty plasma with infinite electrical
conductivity and also with the quasineutrality condition
$n_i\approx n_e+Zn_d$ we can write down the single fluid momentum
equation of the bulk fluid as
  \begin{equation}
  \rho_d\frac{d {\bf v}}{dt} = {{\bf J}\times {\bf B}}+\eta
  \bigtriangledown^{2} {\bf v}+ \left(\xi+\frac{\eta}{3}\right)\nabla (\nabla
  \cdot{\bf v}).
 \label{bdfe}
 \end{equation}
  In the above equation apart from the viscous force on the righthand side
  there is also ${\bf J}\times{\bf B}$ force and to find the
  evolution of the magnetic field we need to find the electric field.
  For this first we add equations (\ref{eem}), (\ref{iem})
  and then use quasineutrality condition ($n_i\approx n_e+Zn_d$) and
  the expression for  current density to get the following form of the
  generalized Ohm's law
  \begin{equation}
   {\bf E}=-{\bf v}\times {\bf B}-\frac{{\bf J}\times {\bf B}}{Zen_d}.
   \label{gol}
   \end{equation}
  Taking curl of the above equation and using Faraday's law ($\nabla
\times {\bf E}
  =- \partial {\bf B}/\partial t$), the time evolution of magnetic field
for the
  bulk dusty plasma can be obtained as,
   \begin{equation}
    \frac{\partial {\bf B}}{\partial t} = \nabla\times({\bf v}\times {\bf
B})+
    \frac{\nabla\times({\bf J}\times {\bf B})}{Zen_d},
    \label{mfe}
    \end{equation}
    where the first term in the right hand side is the convective term and the second one is the Hall term.
    In the limit of large magnetic Reynold's number that is being
    considered here, the magnetic field lines can be assumed to be
    frozen in the dusty plasma and convected with the plasma fluid
    flow when the contributions of the Hall term are neglected.
    The ratio of the Hall to the convection term can be estimated as
    $\sim v_A/L \omega_{cd} \sim \delta_d/L$,
      where
    $\omega_{cd}(=Ze B_0/m_d)$ and
    $\delta_d=v_A/\omega_{cd}=c/\omega_{pd}$ are the dust cyclotron
    frequency and dust skin depth, and $L$ and $v_A(=B_0/\sqrt{\mu_0
    \rho_{d0}})$ are the characteristic length and dust Alfven
    velocity of the system. For waves with scale length $L
    \gg \delta_d$, the Hall term can be neglected.

    A strongly correlated dusty plasma system can be considered as a
    viscoelastic
    medium. In such a medium normal fluid like equations are
    modified due to the growing correlation between dust
    particles. Normal fluid viscosity coefficient in a
    viscoelastic medium becomes viscoelastic operator
    as described in detail in Frenkel's book \cite{fren}.
    We follow the same procedure and write the generalized
    equation of motion of dust fluid in a viscoelastic medium
    as
    \begin{equation}
    \left(1+ \tau \fder{}{t}\right)\left[\rho_d \frac{d {\bf v}}{dt}
    - {\bf J}\times {\bf B} \right ] = \eta \bigtriangledown^{2} {\bf v}
    + \left(\xi+\frac{\eta}{3}\right)\nabla (\nabla \cdot {\bf v})
    \label{gmhe}
    \end{equation}
    where $\tau$ is the relaxation time of the
    medium \cite{fren, sora}.
    Eq.(\ref{gmhe}) can be considered as the
    generalized magnetohydrodynamic equation that contains
    viscoelastic effects. From the above equation it is clear that
    in the absence of viscoelastic effect the equation is simply
    Navier Stokes equation where kinetic pressure is replaced by
    the magnetic pressure. Therefore
    the limit $\omega \tau \ll 1$,  for which
    equation (8) reduces to the standard magnetohydrodynamic equation
    describing magnetized plasmas can also be termed as hydrodynamic
    limit in analogy with the Navier-Stokes like
    equation mentioned before. Equations (\ref{al}), (\ref{mfe})  and
    (\ref{gmhe})
     are magnetohydrodynamic equations
     describing low frequency phenomena in a strongly coupled,
    cold magnetized dusty plasma. Although derived in the context of a
magnetized dusty plasma, these  equations have a general appeal and can
be utilized for investigations of  strongly coupled magnetized fluids such as those
occurring in astrophysical systems.
    \section {Linear stability analysis}
     Before going to the stability analysis it is useful to explain the
     equilibrium. For simplicity we have assumed that in equilibrium
     plasma  is
     homogeneous. The homogeneous plasma is described by the constant
     variables
     $\rho_d=\rho_0$, ${\bf v} = 0,$ ${\bf B} = B_0 \hat{z}$.
     With the equilibrium  mentioned above  we perturbed the system with a
     small
     amplitude perturbations i.e. ${\bf v}= 0+ {\bf v}_1 ({\bf r},t)$,
     ${\bf B}={\bf B}_0+{\bf B}_1({\bf r},t)$ and  ${\bf J}=0+{\bf J}_1
     ({\bf r},t)$
     where all the variables with subscript one are perturbations.
     Linearizing  Eqs. (\ref{al}), (\ref{mfe})  and (\ref{gmhe})
     around the equilibrium mentioned above we have
     \bee
      \nabla \times {\bf B}_1 = \mu_0 {\bf J}_1
      \label{l1},
     \ene
     \bee
     \fpar{{\bf B}_1}{t} = \nabla \times ({\bf v}_1\times{\bf
     B}_0)
     \label{l2},
    \ene
    \bee
    \left(1+ \tau \fpar{}{t}\right)\left[\rho_0 \fpar{{\bf v}_1}{t}
    - {\bf J}_1\times {\bf B}_0 \right ] = \eta \bigtriangledown^{2} {\bf
    v}_1
    + \left(\xi+\frac{\eta}{3}\right)\nabla (\nabla \cdot {\bf v}_1)
    \label{l3}.
    \ene

    We consider that a wave is propagating making an angle $\theta$
    with unperturbed magnetic field ${\bf B}_0$ i.e, wave vector
    ${\bf k}$ and ${\bf B}$ are in the same plane with wave
    vector ${\bf k} = k_{x} \hat x+k_{z} \hat z $. Since the above
    equations are linear we can  Fourier transform these equations
    assuming the solutions for the perturbed variables are in the form
    $\sim \exp[-i(\omega t
    -{\bf k}\cdot {\bf r})]$. Here $\omega$ is the frequency and ${\bf k}$
    is the
    wave vector of the mode under consideration. Substituting
    perturbed solutions in Eqs. (\ref{l1}) - (\ref{l3}) we find
    \bee
    {\bf k} \times {\bf B}_{1}=-i\mu_{0}{\bf J}_{1},
    \label{ft1}
    \ene
    \bee
    (1- i\omega \tau)[-i\omega \rho_{0}{\bf v}_{1}-{\bf J}_{1} \times {\bf
     B}_{0}]=
    -\eta k^{2} {\bf v}_{1}-\left(\xi + \frac{\eta}{3}\right) {\bf k}({\bf
     k}\cdot{\bf v}_{1}),
     \label{ft2}
     \ene
    \bee
    \omega {\bf B}_{1}={\bf B}_{0}({\bf k}\cdot{\bf v}_{1})-({\bf B}_0
    \cdot {\bf k})
    {\bf v}_{1}.
    \label{ft3}
    \ene
    In the limit $\omega \tau\gg 1$, it is possible to get a dispersion
    relation that describes purely
    propagating modes without any dissipative damping. This is known as
    the kinetic limit as
    opposed to the hydrodynamic one.
     In the electrostatic limit ${\bf B}_1=0$ with $\omega \tau \gg 1$,
     from  Eq. (\ref{ft2})
     one can find both
    compressional mode with $\bf{k}\cdot\bf{v}_{1}\neq 0$ and shear
    mode with ${\bf k}
    \times \bf {v}_{1} \neq 0$.  The velocity of the shear wave and the
     compressional
    wave are found to be $v_{sh}^{2}=\eta/\tau \rho_0$
    and $v_{c}^{2}=(\xi+\frac{4}{3}\eta)/\tau\rho_0$ as investigated
before \cite{pkaw}.
    Eliminating ${\bf B}_1$ and ${\bf J}_1$ from above three equations
(\ref{ft1})-(\ref{ft3})
    we have equations for ${\bf v}_1$ as
     \begin{equation}
      [\omega^{2}-v_{A}^{2}k_{z}^{2}-v_{sh}^{2}k^{2}] {\bf v}_{1} +
[(v_{sh}^{2}-v_c^2) {\bf k}
       -{\hat x}k_x v_{A}^{2}]({\bf k}\cdot {\bf v}_{1})
       +k_z v_{A}({\bf v}_A \cdot {\bf v}_{1})
       {\bf k} = 0.
       \label{ve}
      \end{equation}
       To find the dispersion relation we have taken two different kinds of
       polarization for the velocity vector ${\bf v}_1$. First let us take
       ${\bf v}_{1} = v_{1x} \hat x + v_{1z}\hat z$ which means
       the velocity vector is polarized in the ($x-z$) plane i.e. in
       the plane where the propagation vector lies.
       From Eq.(\ref{ve}), considering $x$ and $z$ components
       the dispersion equation in matrix form can be written as,
      \begin{equation}
      \left
       (\begin{array}{lr}
       \omega^{2}-v_{A}^{2}k^{2}-v_{sh}^{2}k_{z}^{2}-v_{c}^{2}k_{x}^{2} &
        -(v_{c}^{2}-v_{sh}^{2})k_{x}k_{z}\\
        -(v_{c}^{2}-v_{sh}^{2})k_{x}k_{z} &
        \omega^{2}-v_{sh}^{2}k_{x}^{2}-v_{c}^{2}k_{z}^{2}\\
         \end{array}\right)
         \left(\begin{array}{c}v_{1x} \\ v_{1z}\\
         \end{array}\right)=0.
         \label{mt}
         \end{equation}
         The dispersion relation can be obtained equating the determinant
         of the matrix to zero
         which is given by
         \begin{equation}
          \frac{\omega^{2}}{k^{2}} =
          \frac{1}{2}(v_{A}^{2}+v_{c}^{2}+v_{sh}^{2})\pm
          \frac{1}{2}\left[v_{A}^{4}+(v_{c}^{2}-v_{sh}^{2})^{2}-2v_{A}^{2}(v_{c}^{2}-
          v_{sh}^{2})\cos2\theta\right]^\frac{1}{2}
          \label{dp}
          \end{equation}

          where $ \cos\theta = k_z/k $. In the limit of both $\xi$ and $\eta$
          going to zero, Eq.(\ref{dp}) reduces to the pure compressional
          Alfven
          wave propagating in a cold plasma that is partly longitudinal and
          partly transverse. For $\theta = 0$, we obtain
          \begin{eqnarray}
           {\omega^2} &=& k^2(v_{sh}^2 + v_A^2) ~~{\rm{for}}~~ v_{1z}=0
            \nonumber\\
           &=& k^2 v_c^2 ~~~~~~~~~~~~{\rm{for}}~~v_{1x} = 0
          \end{eqnarray}
           with the two modes being transverse and  longitudinal
           respectively.
           When the direction of propagation perpendicular to the
           unperturbed magnetic field i.e.
           $\theta =\pi/2 $, then we have
           \begin{eqnarray}
           {\omega^2} &=& k^2(v_c^2+ v_A^2) ~~{\rm{for}} ~~ v_{1z}=0
\nonumber\\
            &=& k^2 v_{sh}^2  ~~~~~~~~~{\rm{for}}~~v_{1x} = 0
           \end{eqnarray}
           The transverse component in this case is a purely mechanical
shear mode independent
           of magnetic field since the Lorentz force vanishes in this case.
           The longitudinal component depends on the magnetic pressure as
well as pressure due to
           viscous forces.
           In the general case when the propagation is oblique with respect
           to the magnetic field we get mixed modes that are partially
           transverse and partially longitudinal type with the
polarization in
           the plane generated by the magnetic field and the propagation
           direction.

           Next, we consider the velocity perturbation
           perpendicular to the direction of propagation vector i.e. ${\bf v}_1= v_{1y} \hat y $.
            From Eq.
           (\ref{ve})we have
           \begin{equation}
            (\omega^{2}-v_{A}^{2}k_{z}^{2}-v_{sh}^{2}k^{2})v_{1y}=0.
            \label{tm1}
           \end{equation}
            For $v_{1y}\neq 0$, a transverse mode propagates in the $x-z$
             plane with phase
            velocity
            \begin{equation}
             v_{p}=\frac{\omega}{k}=\sqrt{(v_{A}^{2}\cos^{2}\theta+v_{sh}^{2})}
            \label{tm2}
            \end{equation}
           When $\theta = 0$ i.e, when the tranverse shear wave is propagating along
            the unperturbed magnetic field(${\bf B}_{0}$), the phase velocity becomes
            \begin{equation}
             v_{p}=\frac{\omega}{k}=\sqrt{(v_{A}^{2}+v_{sh}^{2})}
            \label{final}
            \end{equation}
     In the absence of magnetic field, Eq.(\ref{final}) reduces to the
     dispersion relation for a purely elastic mode obtained in
     Ref. \cite{pkaw}.
             In the magnetohydrodynamic limit (pure fluid with $\eta=0$),
             the above mode
             reduces to the well known shear Alfven wave.
The dispersion relations
for circularly polarized transverse shear Alfven waves have been derived earlier.
In the very low frequency limit $\omega <<\omega_{cd}$, such dispersion relations
reduce to the linearly polarized waves described by Eq.(\ref{final}).

            \section{Conclusions}
            A generalized magnetohydrodynamic equation describing a strongly
            coupled, magnetized, cold dusty plasma has been set up.  The
            equation is utilized to derive the dispersion relation that
            describes coupling between modes that arise due to magnetic and
            viscoelastic stresses. In analogy with the magnetoacoustic modes
            that are sound-like modes that propagate in a magnetized fluid,
            the compressional and shear modes that propagate in a magnetized
            elastic fluid  can be termed as  `magnetoelastic
            modes'.

From eq. (22) describing the transverse shear waves,  after using the
appropriate expressions\cite{pkaw,berk} for $\eta$ and $\tau$, we obtain:
$$ \omega^2 = k^2a_d^2(\frac{\lambda_{Dd}^2}{a_d^2}\omega_{pd}^2 f(\Gamma)+ \frac{c^2}{a_d^2}\frac{\omega_{cd}^2}{\omega_{pd}^2}) $$
with
$$f(\Gamma) = \left[1-\gamma_d\left(1+\frac{u}{3}+\frac{\Gamma}{9}
\frac{\partial u}{\partial \Gamma}+\frac{4}{15}u(\Gamma)\right)\right ] $$
where $a_d$ is the Wigner-Seitz radius, $\gamma_d$ is the adiabaticity constant, and
 $u(\Gamma)$ is the excess internal energy of the system.
In the limit
of   $1\leq \Gamma \leq 200$, $u(\Gamma)=-0.9\Gamma+0.95\Gamma^{1/4}+0.19\Gamma^{-1/4}-0.81$.  In obtaining the dispersion relations for
the magnetoelastic modes, we have neglected the effects of dust-neutral collisions.
The effect of dust-neutral collisions, that are important in many experimental situations
can be incorporated by replacing $\omega^2$ by $\omega(\omega+i\nu_{dn})$, where $\nu_{dn}$
is the dust-neutral collision frequency.  Collisional effects can thus be considered to be negligible\cite{kalman} when the following condition holds
$${\bf{\frac{\nu_{dn}}{\omega_{pd}} << k\lambda_{Dd} \sqrt{f(\Gamma)}.}}$$
For  $\lambda_{Dd}\leq a_d$ and for $ka_d \approx 0.1$, collisions can be neglected.
Since the dust-neutral collision frequency is proportional to neutral
gas pressure, experimentally, the modes can be observed at low neutral gas pressures.

The thermal contribution from electrons, ions and dust particles described by
a total plasma pressure  $p$  is known to lead to
magnetosonic dust modes\cite{verheest}
with the effective pressure given by $p+B^2/\mu_0$
where the comparison is between plasma pressure and magnetic pressure terms.  In the present work,
we have considered only a comparison between mechanical stresses and magnetic stresses without considering the plasma pressure terms.  For
transverse type stresses,  there is no contribution from the plasma pressure terms.

The presence of magnetic fields in a dusty plasma can alter the currents on the dust surface
thereby changing the nature of the dust charging mechanisms.  Theoretical
estimates\cite{sato} have shown that the value of dust charge
in  magnetic fields depends on the size of the dust particle relative to the ion and
electron gyro-radii and for strong fields, the dust charge
can be substantially larger than the values in the absence of
or weak magnetic fields.

For the magnetic and elastic effects to be of comparable importance
 the values of
magnetic field strength should be such  as to satisfy
$$ \frac{\lambda_{Dd}^2\omega_{pd}^2}{c^2}f(\Gamma) \approx \frac{\omega_{cd}^2}{\omega_{pd}^2}$$
This leads to a  dust larmor radius  $\rho_{Ld}$ given by
$\rho_{Ld}\approx {c}/{\omega_{pd} \sqrt{f(\gamma)}}$.
This gives a region of parameter space where the density, magnetic field, temperature
and the coupling parameter $\Gamma$ should satisfy
$ n_0 T_0 f(\Gamma) \approx B^2/\mu_0 $.
The value of dust charge enters  the above condition on different parameters through the
value of $\Gamma$.  Any dependence of the dust charge on the value of the magnetic
field strength should manifest through the value of $\Gamma$ for which no
explicit relation is known yet.
In strongly coupled laboratory plasmas\cite{kono, pintu}, where the typical dust densities and temperatures
are $\sim 10^5$cm$^{-3}$, $3-5$ eV, and  $\Gamma$ in the range $1<\Gamma<50$,
the values of magnetic field where the
magneto-elastic effects are important  will lead to large dust Larmor radii and
 almost unmagnetized dust grains.
For high density  astrophysical plasmas such as those occurring in white dwarfs
and neutron stars where there exist a wide range of magnetic field strengths,
the combined action of Hooke's elastic and Lorentz magnetic force has been suggested\cite{bastrukov}
to consistently interpret the detected Quasi-periodic oscillations.
In such scenario, the generalized magnetohydrodynamic equations incorporating both
viscoelastic and magnetic contributions can be considered as an appropriate model for studying various wave and instabilities.
\bibliography{pp-mem-mod}
\end{document}